\documentclass[%
reprint,
superscriptaddress,
nobibnotes,
 amsmath,amssymb,
 aps,
prb,
footinbib,
]{revtex4-2}
\usepackage{graphicx,subfigure}
\usepackage{epsfig}
\usepackage{bm}				
\usepackage{dcolumn}
\usepackage{color,xcolor}
\usepackage{float}
\usepackage[colorlinks,urlcolor=blue,citecolor=blue,linkcolor=magenta]{hyperref}
\makeatletter
\newcommand*{\rom}[1]{\expandafter\@slowromancap\romannumeral #1@}
\makeatother
\usepackage{lipsum}
\usepackage{subfigure}
\usepackage{dsfont}
\usepackage{enumitem}
\usepackage{tikz,verbatim}
\usetikzlibrary{quantikz}
\usepackage{amsmath,amsfonts,amssymb}
\usepackage{amsthm}
\usepackage{leftidx}
\usepackage{hyperref}
\usepackage{float}
\usepackage[normalem]{ulem}
\usepackage{dcolumn}
\usepackage{bm}
\usepackage{epstopdf}
\usepackage{epsfig}
\usepackage{mathdots}

\newcommand\scalemath[2]{\scalebox{#1}{\mbox{\ensuremath{\displaystyle #2}}}}

\begin{document}

\title{NonHermitian Topological Phases in a Hermitian Modified Bosonic Kitaev Chain}
\author{Raditya Weda Bomantara}
	\email{Raditya.Bomantara@kfupm.edu.sa}
	\affiliation{%
		Department of Physics, Interdisciplinary Research Center for Intelligent Secure Systems, King Fahd University of Petroleum and Minerals, 31261 Dhahran, Saudi Arabia
	}
 \author{Ibsal Assi}
    \affiliation{Department of Physics and Physical Oceanography,
Memorial University of Newfoundland and Labrador,St. John’s, Newfoundland \& Labrador, Canada A1B 3X7}
\author{J. P. F. LeBlanc}
\affiliation{Department of Physics and Physical Oceanography,
Memorial University of Newfoundland and Labrador,St. John’s, Newfoundland \& Labrador, Canada A1B 3X7}
\affiliation{Compute Everything Technologies Ltd., St. John's, Newfoundland \& Labrador, Canada}
    \author{Michael Vogl}
	\affiliation{%
		Department of Physics, Interdisciplinary Research Center for Intelligent Secure Systems, King Fahd University of Petroleum and Minerals, 31261 Dhahran, Saudi Arabia
	}
    \date{\today}


\maketitle

{\bf We present a modification to the bosonic Kitaev chain that, despite being Hermitian, supports both nonHermitian skin effect and nontrivial topological edge modes in its excitation Hamiltonian. We establish an exact mapping between the excitation Hamiltonian of our system and a nonHermitian Su-Schrieffer-Heeger (SSH) model, which allows for a completely analytical characterization of its topology. In particular, topological phase transition points separating a topologically trivial and nontrivial regime were identified analytically by the appropriate winding number invariant and the presence of zero energy modes. Similarly to the regular bosonic Kitaev chain, the nonHermitian skin effect and some (but not all) topological edge modes are quickly destroyed at nonzero bosonic onsite potential (harmonic oscillator frequency). Remarkably, however, disorder partially recovers some of these features. This work thus demonstrates the potential of a modified bosonic Kitaev chain as a platform to generate rich nonHermitian topological phenomena from a completely Hermitian system's perspective. Lastly, we suggest a possible experimental realization of the model, which could allow for total control over the parameter space.}

\section{Introduction} 
\label{intro}

The Kitaev chain is a paradigmatic yet simple toy model of a topological superconductor that supports long-sought-after Majorana modes at the system's edges \cite{Kitaev01}. The model describes a tight-binding one-dimensional (1D) fermionic lattice with $p$-wave superconductivity. In particular, $p$-wave superconductivity here gives rise to several non-conserving terms in the Hamiltonian that simultaneously add or subtract two fermions at neighboring lattice sites. Such a term is crucial in fractionalizing fermions into their Majorana fermionic components. In the topologically nontrivial regime, a pair of Majorana fermions at the edges of the lattice is effectively uncoupled and serves as zero energy excitations of the system, i.e., Majorana zero modes. Majorana zero modes are desirable because they can be used to encode and process quantum information in a topologically protected manner \cite{Nayak08,Lahtinen17}. Consequently, over the last decade, Majorana zero modes and their quantum computing applications have been the subject of extensive theoretical studies \cite{Kraus13,Alicea11,Heck12,Vijay16, Zheng16,Litinski17,Plugge17,Karzig17,Bomantara18,Bomantara18b,Bomantara20,Bomantara20b}. Experimentally, the detection of Majorana zero modes in superconducting systems remains of ongoing interest \cite{Mourik12,Nichele17,Lutchyn18}.    

While the original Kitaev chain is fermionic by construction, recent studies have considered a bosonic variation of the Kitaev chain \cite{Mcdonald18, Ugh24, Slim24, Busnaina24, He25}. First proposed in Ref.~\cite{Mcdonald18}, the so-called bosonic Kitaev chain (BKC) is described by the second-quantized Hamiltonian that takes on the same mathematical form as the fermionic Kitaev chain, but with all fermionic operators replaced by their bosonic counterparts. Due to the difference in commutation relations between bosonic and fermionic operators, the two models yield significantly different physics. For instance, while the number non-conserving term fractionalizes the fermionic operators into mutually anticommuting Majorana fermions, it instead fractionalizes bosonic operators into mutually commuting quadrature operators. Moreover, the equations of motion governing the dynamics of quadrature operators in the BKC differ significantly from those of Majorana operators in the fermionic counterpart. In particular, the former could effectively be described by a nonHermitian Schr\"{o}dinger equation, which potentially yields a plethora of interesting phenomena.

It is worth noting that nonHermitian systems have recently emerged as an active field of research \cite{Rudner09,Shen18,Yao18,Kunst18,Gong18,Li18,Lee19,Yokomizo19,Yang20,Ji25,Alvarez18,Okuma20,Zhang20,Song19,Lee19b,Ashida20,Coulais20,Bergholtz21,Okuma23} because they play host to phenomena with no Hermitian counterparts. A popular example is the nonHermitian skin effect (NHSE), where all bulk energy eigenstates are localized close to the system's edges \cite{Yao18}. The NHSE is also accompanied by a sensitivity of the energy spectrum to the choice of boundary conditions. This observation led to a reformulation of the bulk-boundary correspondence for studies of nonHermitian topological phases \cite{Kunst18,Yokomizo19,Yang20,Ji25}. Physically, nonHermitian systems are typically expected to arise as effective descriptions of open quantum system \cite{Daley14} or appear as a description of a classical system such as an acoustic lattice \cite{Zhang21}, light propagating through a waveguide lattice \cite{Weimann17}, or electrical circuits \cite{Helbig20,Zou21,Liu21,Liu23}. The BKC and its variations yield another means to generate nonHermitian physics that, unlike other typical realizations of nonHermitian systems, is fully quantum.

In this work, we investigate a variation of the BKC, which incorporates elements from the Su-Schrieffer-Heeger (SSH) model, the latter of which is well-known for its intuitive formation of topological edge modes \cite{Su80}. The primary motivation for considering the proposed model is to uncover phenomena that arise from the interplay between the effective nonHermitian description of the BKC and the topological characteristics of the SSH model. Our analysis focuses on the effective nonHermitian excitation Hamiltonian (see below for its precise definition) of the proposed system, particularly on its spectral and topological properties. Our key findings include an exact mapping of the system's excitation Hamiltonian to an analytically solvable nonHermitian SSH Hamiltonian. This mapping is valid for arbitrary system parameters. Moreover, we observe the formation of two types of topological edge modes, their coexistence with the NHSE, and their robustness against spatial disorder.

This paper is organized as follows. In Sec.~\ref{overview}, we set up notation and definitions, then apply them in a mathematical review of the BKC. We further analytically and numerically demonstrate the formation of the NHSE in the BKC. In Sec.~\ref{modified}, we introduce our proposed model that combines the BKC with the SSH model. In Sec.~\ref{NHSE}, we numerically verify the presence of the NHSE in the proposed model by plotting the energy excitation spectrum under different boundary conditions and the spatial profiles of all eigenstates of the excitation Hamiltonian. In Sec.~\ref{topology}, we analytically and numerically uncover the presence of topological edge modes and their underlying topological invariants. This approach allows for a complete characterization of the topology in the proposed model. In Sec.~\ref{disorder}, we investigate the impact of spatial disorder on the established NHSE and topological edge modes. In Sec. \ref{sec:exp_real}, we discuss a possible experimental realization of the model. Finally, Sec.~\ref{conc} summarizes the paper and presents potential future research directions.

\section{Overview of BKC}
\label{overview}

The BKC is described by the Hamiltonian \cite{Mcdonald18}
	\begin{equation}
	H=\sum_{j=1}^{N-1} \left(J a_j^\dagger a_{j+1} + \Delta a_j^\dagger a_{j+1}^\dagger +h.c. \right) +\sum_{j=1}^N \omega a_j^\dagger a_j \;, \label{SSH_boson}
	\end{equation}	
where $N$ is the chain length, $a$ is the bosonic operator, $J$ and $\Delta$ are the hopping and the bosonic analogs of pairing amplitudes, respectively, and $\omega$ is the bosonic frequency. We stress that the last onsite energy term $\omega$ was added to the model discussed in \cite{Mcdonald18} for slight generalization. Its effect on the BKC has also been discussed in great detail in \cite{Ugh24}.

Next, to cast the Hamiltonian into a form that more readily reveals topological features, we recall that a bosonic operator could be written as a linear combination of Hermitian position and momentum operators (following the discussion by \cite{Mcdonald18} we will call them quadrature operators). In dimensionless units the expressions read $a_j= \frac{x_j + \mathrm{i} p_j}{\sqrt{2}}$. This reformulation in terms of $p_i$ and $x_i$ is useful because it more closely resembles a Majorana description from the fermionic case (this hints at the fact that these variables might provide topological insights) and it readily exposes connections to classical physical systems, which could suggest possible experimental realizations. As pointed out in \cite{Mcdonald18}, the hopping $J$ needs to have a phase contribution to obtain a topologically nontrivial regime. Therefore, in the following, to have an interesting parameter regime, we will assume that $J$ and $\Delta$ are purely imaginary with $J=\mathrm{i} J_0$ and $\Delta= \mathrm{i} \Delta_0$, where $J_0$ and $\Delta_0$ are the (real) hopping and pairing magnitudes, respectively. Equation~(\ref{SSH_boson}) can then be written as
\begin{eqnarray}
    H&=&\sum_{j=1}^{N-1} \left[-(J_0-\Delta_0) x_j p_{j+1} + (J_0+\Delta_0) p_j x_{j+1} \right] \nonumber \\
    && +\sum_{j=1}^N \frac{\omega}{2} (x_j^2 + p_j^2) .
\end{eqnarray}
As expected, choosing variables $x_i$ and $p_i$ directly yields insights into the model's inner workings. That is, we find that for special parameter values $J_0=\Delta_0$ and $\omega=0$, $p_1$ and $x_{N}$ commute with the Hamiltonian and thus correspond exactly to what we will define as zero modes. To quantify the fate of these zero modes at more general parameter values, we consider linear combinations $q=\alpha_1 x_1 +\cdots +\alpha_n x_n +\beta_1 p_1 + \cdots +\beta_n p_n $ and solve the ``eigenvalue" equation
\begin{equation}
    [H,q] = \epsilon q , \label{excmode}
\end{equation}
where $\epsilon$ is the excitation energy associated with the eigenmode $q$. Intuitively, $q$ maps a reference energy eigenstate $|E\rangle$ of $H$ with energy $E$ to another energy eigenstate $q|E\rangle \propto |E+\epsilon\rangle$ of $H$ with energy $E+\epsilon$. Thus, $q$ acts like a ladder operator, raising the energy by $\epsilon$. Zero modes are then defined as solutions $q$ of Eq. \eqref{excmode} with excitation energy $\epsilon=0$.     

Equation~(\ref{excmode}) can be turned into a matrix equation by first evaluating
\begin{equation}
   \scalemath{0.95}{ \begin{aligned}
    [H,x_1] &= -\mathrm{i}(J_0 +\Delta_0) x_2 -\mathrm{i} \omega p_1   \\
    \left[H,p_1\right] &= -\mathrm{i} (J_0 -\Delta_0) p_2 +\mathrm{i} \omega x_1   \\
    \left[H,x_{j>1}\right] &= \mathrm{i} (J_0 -\Delta_0) x_{j-1} -\mathrm{i} (J_0 +\Delta_0) x_{j+1} -\mathrm{i} \omega p_j  \\
    \left[H,p_{j>1}\right] &= -\mathrm{i} (J_0 -\Delta_0) p_{j+1} +\mathrm{i} (J_0 +\Delta_0) p_{j-1} +\mathrm{i} \omega x_j
     \end{aligned}}
\end{equation}
where we have used $[x_j,p_\ell]=\mathrm{i} \delta_{j,k} $ ($\hbar=1$ units are used throughout). We note that the set $\left(x_1,\cdots,x_n,p_1,\cdots,p_n\right)$ is closed under the commutator $[H,\cdots]$. Moreover, we also note that for a quadratic Hamiltonian (like in our case), classical Hamilton's equations of motion yield the same set of equations, which points to possible classical realizations in terms of unstable coupled oscillators.

Thus, the matrix representation of $[H,\cdots]$ in the basis of $\left(x_1,\cdots,x_n,p_1,\cdots,p_n\right)$ is given as
\begin{eqnarray}
    [H,\cdots] &\hat{=}& \mathcal{H} , \nonumber \\
    \mathcal{H} &=& -\mathrm{i} \sum_{j=1}^{N-1} \left[ \left(J_0 \sigma_0 + \Delta_0 \sigma_z\right) \otimes |j+1\rangle \langle j| \right. \nonumber \\
    && \left.- \left(J_0 \sigma_0 - \Delta_0 \sigma_z\right) \otimes |j\rangle \langle j+1| \right] \nonumber \\
    && +\sum_{j=1}^{N} \omega \sigma_y \otimes |j\rangle \langle j| , \label{excham}
\end{eqnarray}
where $\sigma_j$ for $j=0,x,y,z$ are the $2\times 2$ identity and Pauli matrices acting on the quadrature subspace - i.e. the subspace of $(x,p)$. Moreover, kets $|j\rangle$ are column vectors with 1 as the $j$th element and zero elsewhere, describing lattice space degrees of freedom. The matrix $\mathcal{H}$ shall be referred to as the excitation Hamiltonian throughout the remainder of this paper, and its eigenvalues are termed excitation energies.

Note that the matrix in Eq.~(\ref{excham}) is nonHermitian due to the terms involving $\Delta_0$. Moreover, at $\omega=0$, Eq.~(\ref{excham}) is equivalent to two copies of the well-known Hatano-Nelson model \cite{Hatano96,Hatano97}. One remarkable feature of the Hatano-Nelson model is the sensitivity of its eigenvalue spectrum to boundary conditions. Indeed, the energy excitation spectrum is purely real for open boundary conditions (OBC) since there exists a similarity transformation $A$ that maps the matrix Hamiltonian to a Hermitian model \cite{Okuma23} as demonstrated below
\begin{equation}
    A^{-1}\mathcal{H} A = \mathrm{sgn}(\Delta_0) \sqrt{J_0^2 -\Delta_0^2} \sigma_0 \otimes \left( -\mathrm{i} |j\rangle \langle j+1 | + h.c. \right) , 
\end{equation}
where 
\begin{equation}
    A = \left\lbrace r^{\frac{j}{2}} \frac{\sigma_0 +\sigma_z}{2}+ r^{-\frac{j}{2}} \frac{\sigma_0 -\sigma_z}{2} \right\rbrace \otimes |j\rangle \langle j | .
\end{equation}
We introduced the parameter $r=\frac{\Delta_0+J_0}{\Delta_0-J_0}$ for brevity of notation. It is interesting to observe that $A^{-1}\mathcal{H} A$ is Hermitian as long as $J_0>\Delta_0$. In this parameter range, eigenvalues are purely real and given as $\pm 2 \sqrt{J_0^2 -\Delta_0^2} \cos\left(\frac{2\pi j}{N}\right)$ for $j=0,1,\cdots, N-1$. This result is contrasted by the complementary parameter range  $J_0<\Delta_0$, where $A^{-1}\mathcal{H} A$ is anti-Hermitian with purely imaginary eigenvalues.

\begin{center}
    \begin{figure}
        \centering
        \includegraphics[scale=0.45]{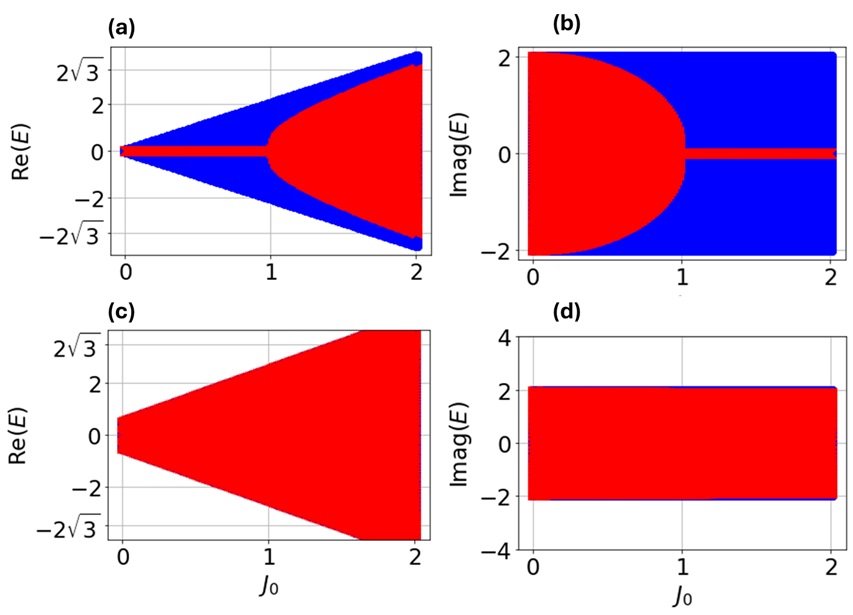}
        \caption{The eigenvalue spectrum of the Hatano-Nelson Hamiltonian under OBC (red) and under PBC (blue). In all panels, the system parameters are set as $\Delta_0=1$, $N=100$, (a,b) $\omega=0$, (c,d) $\omega=0.5$. }
        \label{fig:HNmodel}
    \end{figure}
\end{center}

For periodic boundary conditions (PBC), $\mathcal{H}$  can be written as
\begin{equation}
    \mathcal{H} = (2J_0 \sin(k) \sigma_0 - \mathrm{i} 2\Delta_0 \cos(k) \sigma_z +\omega \sigma_y ) \otimes |k\rangle \langle k| ,
\end{equation}
where $k\in \left(-\pi,\pi \right]$ is the so-called quasi-momentum. Excitation energies are readily found as $2J_0 \sin(k) \pm \sqrt{\omega^2 - 4\Delta_0^2 \cos^2(k)} $. At $\omega=0$, these are generically complex and thus contrast the purely real or imaginary excitation energy solutions in the OBC case. In Fig.~\ref{fig:HNmodel}(a,b), the eigenvalues $E$ of the Hatano-Nelson Hamiltonian (i.e., $\omega=0$ case) for OBC and PBC are plotted side-by-side to highlight their difference. The difference between OBC and PBC eigenvalue solutions indicates the presence of the NHSE, a phenomenon in which all OBC eigenstates are localized at the system's boundaries \cite{Yao18}. Indeed, we have verified that, for representative values of $J_0$ and $\Delta_0$, as long as $\omega=0$, exactly half the eigenstates of $\mathcal{H}$ are localized near lattice site $j=0$, whilst the other half is localized near $j=N-1$. It is important to note that the many edge states we observed here generically do not correspond to zero excitation energy. In this case, the zero modes $p_1$ and $x_N$ we previously identified for parameter values of $J_0=\Delta_0$ and $\omega=0$ do not play a special role; rather, they are just one of many other edge states of the system. 

Finally, in Fig.~\ref{fig:HNmodel}(c,d), we plotted the energy excitations of $\mathcal{H}$ in the case of nonzero $\omega$. Interestingly, the energy excitation spectra for OBC and PBC are now identical, which signifies the absence of the NHSE. To verify, we have checked that no edge states exist in this parameter regime. This breakdown of NHSE already occurs even at a very small value of $\omega$. Consequently, the system's various edge modes (including the zero modes) that exist at $\omega=0$ are not robust. They are easily destroyed by the presence of nonzero $\omega$. For the above reasons, at least in the context of topology and its potential applications, the BKC of Eq.~(\ref{SSH_boson}) is not as attractive as its fermionic counterpart. 

\section{Modified BKC}

\subsection{Model description}
\label{modified}

Next, we introduce a modified version of the BKC that we demonstrate supports richer topology. Our model is based on a combination of the BKC and the paradigmatic SSH model \cite{Su80}. Our Hamiltonian takes the form
\begin{eqnarray}
    H'&=&\sum_{j=1}^{N-1} \left(J_1 a_{A,j}^\dagger a_{B,j} +J_2 a_{B,j}^\dagger a_{A,j+1} + \Delta_1 a_{A,j}^\dagger a_{B,j}^\dagger \right. \nonumber \\
    &&+ \left. \Delta_2 a_{B,j}^\dagger a_{A,j+1}^\dagger +h.c. \right) \nonumber \\
    && + \sum_{j=1}^N \omega \left( a_{A,j}^\dagger a_{A,j} + a_{B,j}^\dagger a_{B,j} \right) ,
\end{eqnarray}
where $A$ and $B$ abstractly represent two sublattices or other degrees of freedom, such as orbital. Parameters $J_1$ and $J_2$ describe intracell and intercell hopping, $\Delta_1$ and $\Delta_2$ are the bosonic analogs of intracell and intercell pairing amplitudes. For simplicity, we shall now set all parameters $J_1$, $J_2$, $\Delta_1$, and $\Delta_2$ as real numbers. In terms of quadrature variables $x_j$ and $p_j$, the Hamiltonian reads
\begin{equation}
    \begin{aligned}
H^{\prime}&= \sum_{j=1}^N\left[\frac{\omega}{2}\left(x_{A, j}^2+p_{A, j}^2+x_{B, j}^2+p_{B, j}^2\right)\right.\\ 
&\left.+(J_1+\Delta_1) x_{A, j} x_{B, j} +(J_1-\Delta_1) p_{A, j} p_{B, j}\right]\\&+\sum_{j=1}^{N-1}\left[(J_2+\Delta_2) x_{B, j} x_{A, j+1}+(J_2-\Delta_2) p_{B, j} p_{A, j+1}\right] .
\end{aligned}
\end{equation}
Corresponding commutators between the Hamiltonian $H^\prime$ and quadrature variables $x_j$ and $p_j$ - much like in the case of the original BKC Hamiltonian - allow us to find an excitation Hamiltonian $\mathcal{H}^\prime$. For completeness, the commutators are listed below
\begin{equation}
    \begin{aligned}
{\left[H^{\prime}, x_{A, j}\right] } & =-\mathrm{i} (J_1-\Delta_1) p_{B, j}-\mathrm{i} (J_2-\Delta_2) p_{B, j-1}-\mathrm{i} \omega p_{A, j} \\
{\left[H^{\prime}, p_{A, j}\right] } & =\mathrm{i} (J_1+\Delta_1) x_{B, j}+\mathrm{i} (J_2+\Delta_2) p_{B, j-1}+\mathrm{i} \omega x_{A, j} \\
{\left[H^{\prime}, x_{B, j}\right] } & =-\mathrm{i} (J_1-\Delta_1) x_{A, j}-\mathrm{i} (J_2-\Delta_2) p_{A, j+1}-\mathrm{i} \omega p_{B, j} \\
{\left[H^{\prime}, p_{B, j}\right] } & =\mathrm{i} (J_1+\Delta_1) x_{A, j}+\mathrm{i} (J_2+\Delta_2) x_{A, j+1}+\mathrm{i} \omega x_{B, j}
\end{aligned}
\end{equation}
Like in the previous section, the excitation Hamiltonian is given as a matrix representation of the commutator $[H',\cdots]$ and written as
\begin{eqnarray}
    \mathcal{H}' &=& \left(J_1 \sigma_y +\mathrm{i} \Delta_1 \sigma_x \right) \tau_x + \left(J_2 \sigma_y + \mathrm{i}\Delta_2 \sigma_x \right) \cos(\hat{k}) \tau_x  \nonumber \\
    &&  +(J_2 \sigma_y +\mathrm{i} \Delta_2 \sigma_x) \sin(\hat{k}) \tau_y +\omega \sigma_y \tau_0 ,
\end{eqnarray}
where $\sigma$'s and $\tau$'s are Pauli matrices acting on the quadrature and sublattice degrees of freedom, respectively, whilst $e^{\pm \mathrm{i} \hat{k}}$ is an operator that shifts lattice sites by one unit.  

\subsection{PBC vs OBC spectrum and NHSE}
\label{NHSE}

\begin{center}
    \begin{figure}
        \centering
        \includegraphics[scale=0.4]{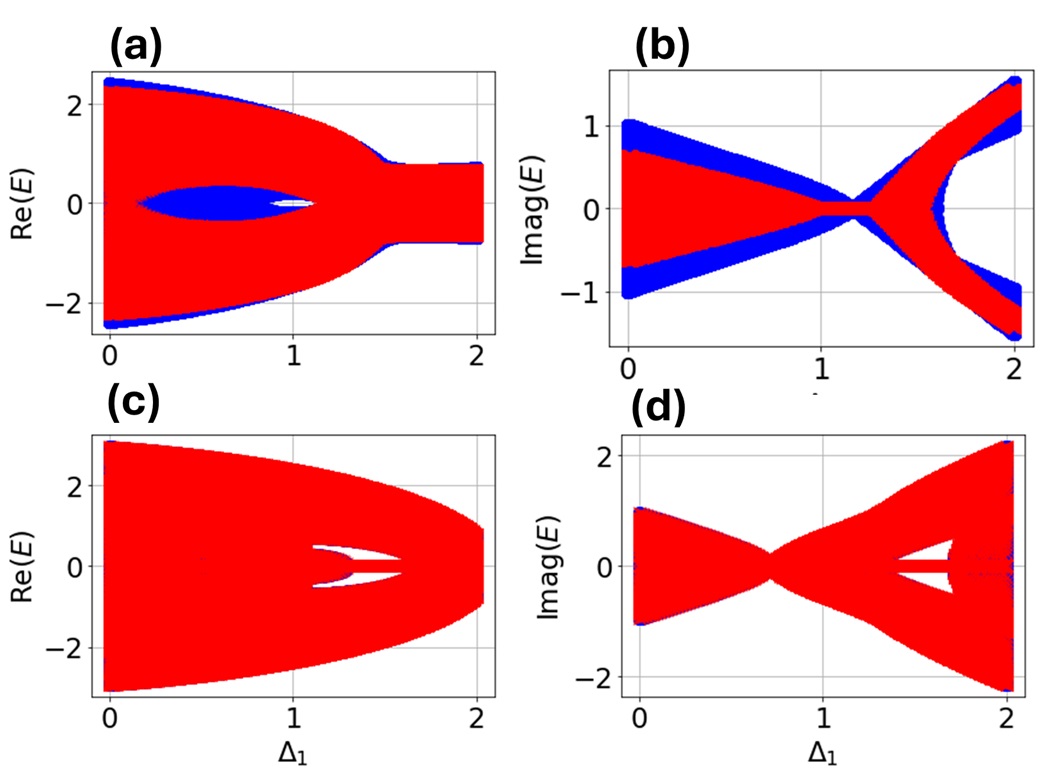}
        \caption{The excitation energy spectrum of $\mathcal{H}'$ under OBC (red) and under PBC (blue) at varying $\Delta_1$. In all panels, the system parameters are set as $J_1=1.4$, $J_2=1.2$, $\Delta_2=1$, $N=100$, (a,b) $\omega=0$, (c,d) $\omega=0.5$.}
        \label{fig:HN2modela}
    \end{figure}
\end{center}

\begin{center}
    \begin{figure}
        \centering
        \includegraphics[scale=0.35]{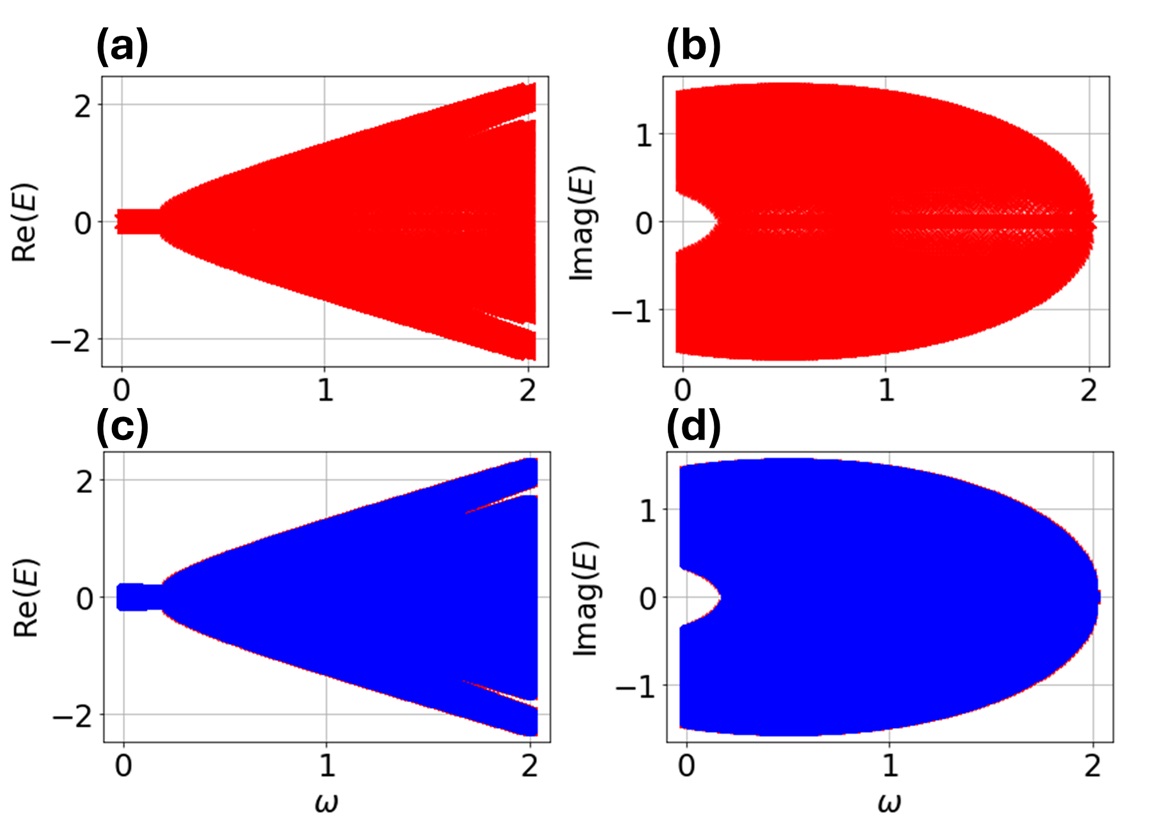}
        \caption{The excitation energy spectrum of $\mathcal{H}'$ under OBC (red) and under PBC (blue) at varying $\omega$. In all panels, the system parameters are set as $J_1=0.4$, $J_2=0.1$, $\Delta_1=1$ $\Delta_2=0.5$, and $N=100$.}
        \label{fig:HN2modelb}
    \end{figure}
\end{center}

We start our investigation by numerically diagonalizing $\mathcal{H}'$ for the case of PBC and OBC. We plot the resulting spectrum as a function of different system parameters to help identify the potential emergence of the NHSE. Our results are presented in Figs.~\ref{fig:HN2modela} and \ref{fig:HN2modelb}. Like the regular BKC model, the energy excitation spectra of $\mathcal{H}'$  for PBC and OBC  differ significantly for $\omega=0$. For nonzero $\omega$, however, no matter how small, both spectra become exactly identical (see Fig.~\ref{fig:HN2modelb}). This observation signifies the absence of NHSE for $\omega\neq 0$.

\begin{center}
    \begin{figure}
        \centering
        \includegraphics[scale=0.43]{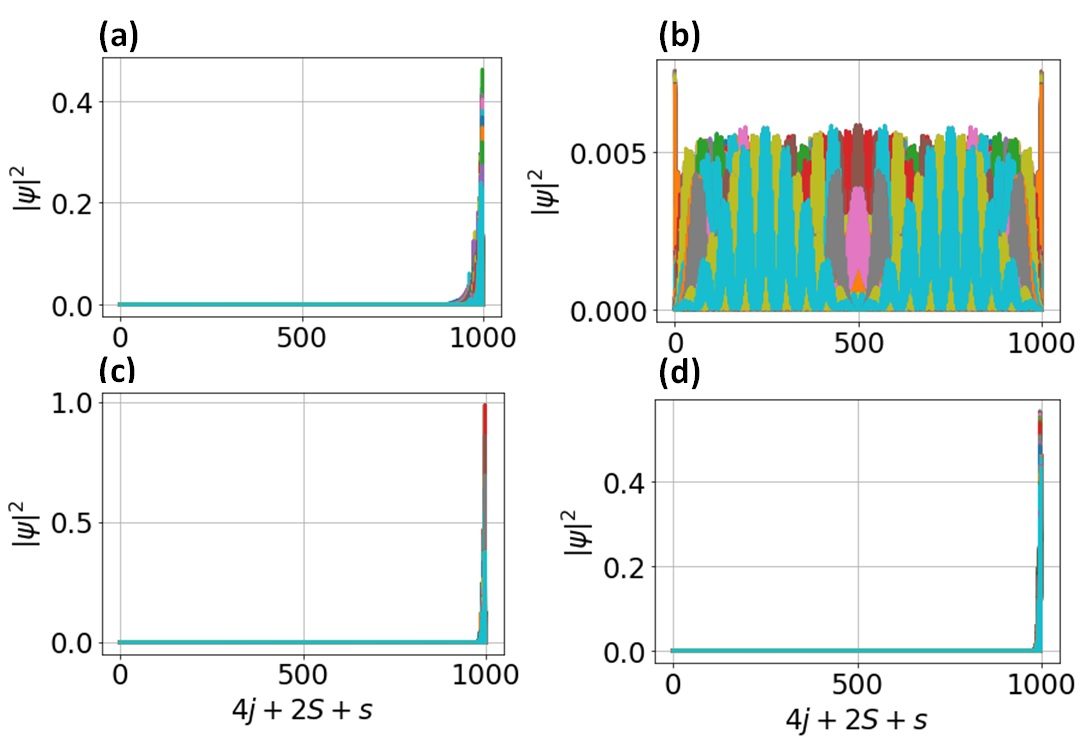}
        \caption{The spatial profiles of all eigenstates of $\mathcal{H}'$ at $N=100$ (see Eq.~(\ref{sprof}) for definitions). System parameters are chosen as (a) $J_1=0.4$, $J_2=0.1$, $\Delta_1=1$, $\Delta_2=0.5$, $\omega=0$, (b) $J_1=0.4$, $J_2=0.1$, $\Delta_1=1$, $\Delta_2=0.5$, $\omega=0.1$, (c) $J_1=0.4$, $J_2=0.1$, $\Delta_1=0.5$, $\Delta_2=1$, $\omega=0$, (d)  $J_1=0.1$, $J_2=0.4$, $\Delta_1=1$, $\Delta_2=0.5$, $\omega=0$.}
        \label{fig:HNwf}
    \end{figure}
\end{center}

Next, we directly verify the presence or absence of NHSE by plotting the spatial profiles of \emph{all} eigenstates of $\mathcal{H}'$. That is, define the occupation probability
\begin{equation}
    |\psi|^2(4j+2S+s)=\langle j,S,s | \psi \rangle , \label{sprof}
\end{equation}
where $|\psi\rangle$ is an eigenstate of $\mathcal{H}'$. Here, $j$ is the site number, $S=0$ ($1$) for sublattice A (B), and $s=0$ ($1$) in the $x$ ($p$) subspace. The combination $4j+S+s$ is chosen so that a unique positive integer is assigned for each tuple of quantum numbers $(j,S,s)$. As demonstrated in Fig.~\ref{fig:HNwf}(a), for $\omega=0$, all eigenstates of $\mathcal{H}'$ are localized at the edges, demonstrating the NHSE. However, for nonzero $\omega$, similarly to the regular BKC, the NHSE is broken, and all eigenstates are delocalized, as shown in Fig.~\ref{fig:HNwf}(b). 


\subsection{Topological edge states}
\label{topology}
Next, we want to do a deeper analysis of topological edge states. Here, we first recall the well-known fact that the fermionic SSH model can be topologically trivial or nontrivial depending on the values of the intracell and intercell hopping amplitudes \cite{Su80}. An interesting feature in the topologically nontrivial regime is that the SSH model supports a pair of robust edge-localized eigenstates at zero energy, i.e., the so-called topological zero modes. It is then interesting to ask if such topological zero modes are also present in our SSH BKC model. Indeed, such states may hide among the non-topological edge-localized eigenstates arising from the NHSE. To systematically uncover the topology of our system, we present our analysis by first setting $J_1$, $J_2$, and $\omega$ to zero, then gradually moving on to the more general cases.

\vspace{0.5cm}
\noindent {\it Case i:} $J_1, J_2, \omega = 0$
\vspace{0.5cm}

\noindent The effective nonHermitian Hamiltonian reduces to $\mathcal{H}'=\mathrm{i} \sigma_x \mathcal{H}_{SSH}$, where
\begin{equation}
    \mathcal{H}_{SSH} = \Delta_1 \tau_x + \Delta_2 \cos(\hat{k}) \tau_x +  \Delta_2 \sin(\hat{k}) \tau_y
\end{equation}
is the regular SSH model. In this case, apart from its energy excitation spectrum being purely imaginary, $\mathcal{H}'$ inherits all physical properties of the SSH model as it essentially comprises of two decoupled copies of $\mathcal{H}_{SSH}$, i.e., due to the tensor product structure with a single $\sigma_x$. In particular, the system is topologically nontrivial (has a nonzero winding number and supports topological zero modes) for $\Delta_2>\Delta_1$, and it is topologically trivial otherwise. Moreover, NHSE is clearly absent for these special parameter choices since the energy spectra are identical under PBC and OBC \cite{Yao18}.

\vspace{0.5cm}
\noindent {\it Case ii:} $J_2,\omega = 0$ and $J_1 \neq 0$
\vspace{0.5cm}

\begin{center}
    \begin{figure}
        \centering
        \includegraphics[scale=0.37]{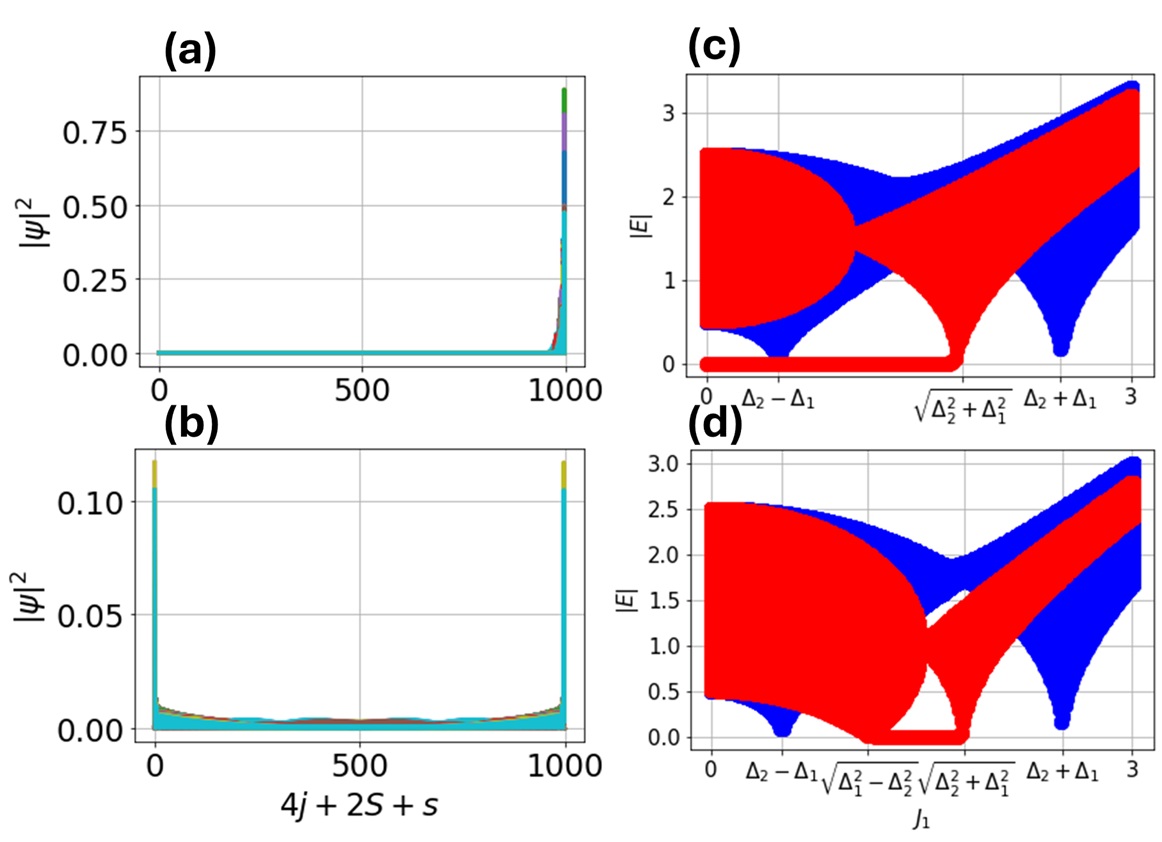}
        \caption{The spatial profiles of all eigenstates of $\mathcal{H}'$ and their energy excitation spectrum at varying $J_1$. In panels (a,b,c), the system parameters are taken as $\Delta_2=1.5$, $\Delta_1=1$, and $N=100$ ($J_1=0.5$ is additionally taken in panels (a,b)). The system parameters in panel (d) are $\Delta_1=1.5$, $\Delta_2=1$, and $N=100$. We set $\omega=0$ in panels (a,c,d) and $\omega=0.05$ in panel (b).}
        \label{fig:spcase1}
    \end{figure}
\end{center}

\noindent By defining the matrix 
\begin{eqnarray}
    A_1' &=& \left\lbrace r_1^{-\frac{j+1}{2}} \frac{\sigma_0 +\sigma_z}{2}\frac{\tau_0+\tau_z}{2} + r_1^{-\frac{j}{2}} \frac{\sigma_0 -\sigma_z}{2}\frac{\tau_0-\tau_z}{2} \right. \nonumber \\ \nonumber \\
    &+& \left. r_1^{\frac{j}{2}} \frac{\sigma_0 +\sigma_z}{2}\frac{\tau_0-\tau_z}{2} + r_1^{\frac{j+1}{2}} \frac{\sigma_0 -\sigma_z}{2}\frac{\tau_0+\tau_z}{2} \right\rbrace \otimes |j\rangle \langle j| \nonumber \\
\end{eqnarray}
where $r_1=\frac{\Delta_1+J_1}{\Delta_1-J_1}$, it can be verified that
\begin{equation}
    A_1'^{-1} \mathcal{H}' A_1' = \mathrm{i} \sigma_x\left[ \tilde{\Delta}_1\tau_x +\Delta_2 \cos(\hat{k})  \tau_x + \Delta_2 \sin(\hat{k}) \tau_y \right] ,  \label{SSHtrans}
\end{equation}
where $\tilde{\Delta}_1=\sqrt{\Delta_1^2-J_1^2}$. That is, $\mathcal{H}'$ is mappable to an SSH model, whose energy excitation spectrum is purely imaginary for $\Delta_1 >J_1$. However, unlike the case of $J_1 = 0$, the structure of the site-dependent similarity matrix $A_1'$ suggests the presence of the NHSE \cite{Yao18}. Indeed, as $r_1^j$ factor in $A_1'$ becomes exponentially very large near one end of the chain, all eigenstates of $A_1'^{-1} \mathcal{H}' A_1'$, which are typically delocalized in nature, becomes extremely localized near one end of the chain, hence establishing NHSE. This observation is indeed confirmed numerically in Fig.~\ref{fig:spcase1}(a), where \emph{all} eigenstates of $\mathcal{H}'$ are sharply localized near the last site. For small but nonzero values of $\omega$, the NHSE is quickly broken, as evidenced in Fig.~\ref{fig:spcase1}(b). Nevertheless, exactly two localized eigenvectors still exist (see Fig.~\ref{fig:spcase1}(b)), which correspond to the genuine topological edge states. 

In the $\omega=0$ case, boundary conditions impact the location of gap closings in parameter space, potentially marking topological phase transitions. Specifically, Eq.~(\ref{SSHtrans}) implies that for OBC, a gap closing occurs at $\Delta_2=\pm \tilde{\Delta}_1$ and $\Delta_2=\pm \mathrm{i} \tilde{\Delta}_1$. These conditions for our original model parameters translate to $J_1^2 =\Delta_1^2 -\Delta_2^2$ and $J_1^2 =\Delta_2^2 +\Delta_1^2$. This observation is contrasted by the case of PBC, where gap closing occurs at $J_1^2=(\Delta_1 \pm \Delta_2)^2 $. In Fig.~\ref{fig:spcase1}(c,d), we further present the plots of $|E|$ vs $J_1$ (we still keep $J_2=0$) to highlight the different closing points between the OBC and PBC cases.

\begin{center}
    \begin{figure}
        \centering
        \includegraphics[scale=0.37]{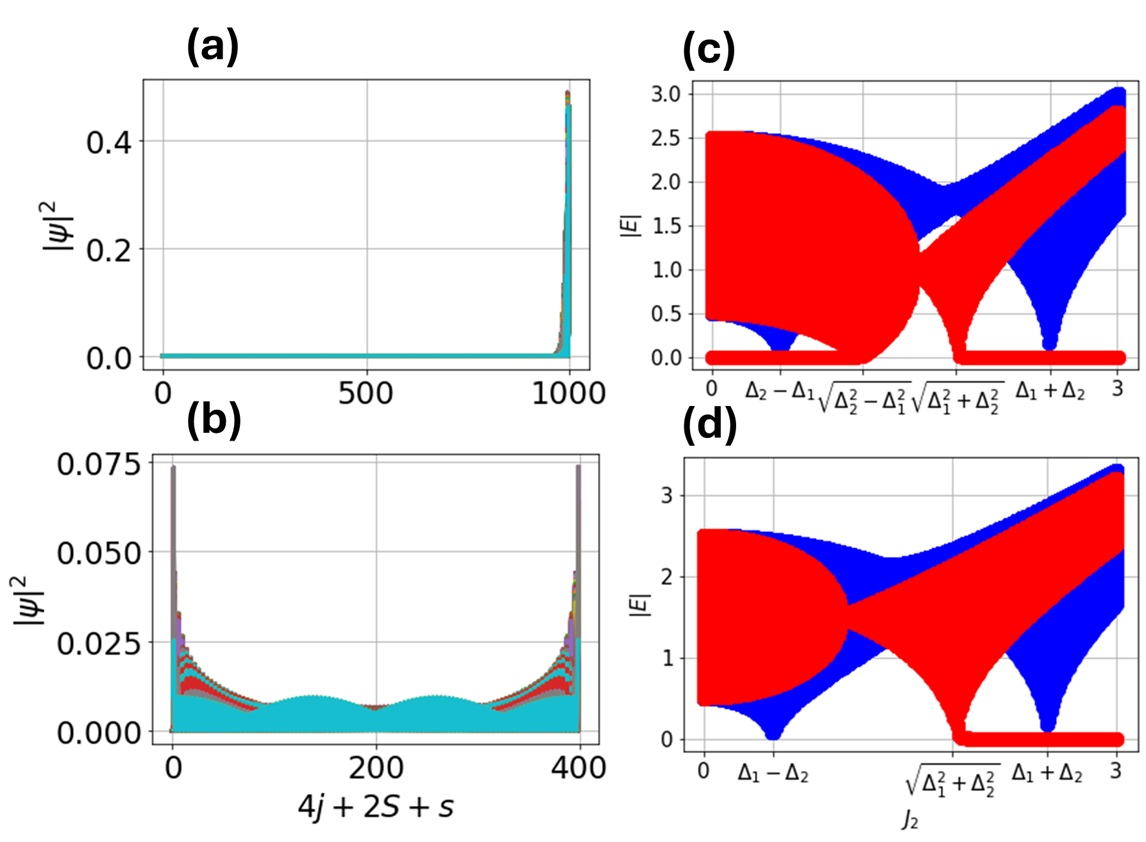}
        \caption{The spatial profiles of all eigenstates of $\mathcal{H}'$ and their energy excitation spectrum at varying $J_2$. In panels (a,b,c), the system parameters are taken as $\Delta_2=1.5$, $\Delta_1=1$, and $N=100$ ($J_2=0.5$ is additionally taken in panels (a,b)). The system parameters in panel (d) are $\Delta_1=1.5$, $\Delta_2=1$, and $N=100$. We set $\omega=0$ in panels (a,c,d) and $\omega=0.05$ in panel (b).}
        \label{fig:spcase2}
    \end{figure}
\end{center}

In Fig.~\ref{fig:spcase1}(c,d), the system is found to be in the topologically nontrivial regime, characterized by the presence of in-gap $|E|=0$ solutions, whenever $\sqrt{\Delta_1^2-\Delta_2^2} <J_1 <\sqrt{\Delta_1^2+\Delta_2^2}$. Additional topological features of the system in this regime can be exposed by calculating the winding number associated with $A_1'^{-1} \mathcal{H}' A_1'$. Specifically, by Fourier transforming $A_1'^{-1} \mathcal{H}' A_1'$, it becomes a block diagonal matrix characterized by the ``quasimomentum" $k\in[-\pi,\pi)$, each subblock of which can be written in the matrix form. 
\begin{eqnarray}
    \tilde{\Delta}_1 \tau_x + \Delta_2 \cos(k) \tau_x + \Delta_2 \sin(k) \tau_y &=& \left( \begin{array}{cc}
        0 & h_- (k) \\
        h_+ (k) & 0
    \end{array} \right) \nonumber \\
\end{eqnarray}
where 
\begin{equation}
h_\pm (k) = \tilde{\Delta}_1 +\Delta_2 \cos(k) \pm \mathrm{i} \Delta_2 \sin(k) .    
\end{equation}
We then define the topological winding number as
\begin{equation}
    w_\pm = \frac{1}{2\pi \mathrm{i}} \oint h_{\pm}^{-1} dh_{\pm} , \label{wn}
\end{equation}
which can be analytically computed as $w_\pm = 0$ (topologically trivial) for $|\tilde{\Delta}_1|>\Delta_2$ and $w_\pm = \pm 1$ (topologically nontrivial) for $|\tilde{\Delta}_1|<\Delta_2$. The latter condition expressed in terms of the original model parameters translates to $\sqrt{\Delta_1^2-\Delta_2^2} <|J_1| <\sqrt{\Delta_1^2+\Delta_2^2}$, which overlaps exactly with the presence of in-gap $|E|=0$ solutions in Fig.~\ref{fig:spcase1}(c,d). This observation proves that such solutions indeed correspond to topological zero modes.

\vspace{0.5cm}
\noindent {\it Case iii:} $J_1,\omega = 0$ and $J_2 \neq 0$
\vspace{0.5cm}

\noindent Similar to case ii), we define another transformation matrix
\begin{eqnarray}
    A_2' &=& \left\lbrace r_2^{-\frac{j}{2}} \frac{\sigma_0 -\sigma_z}{2}\frac{\tau_0+\tau_z}{2} + r_2^{-\frac{j}{2}} \frac{\sigma_0 +\sigma_z}{2}\frac{\tau_0-\tau_z}{2} \right. \nonumber \\ \nonumber \\
    &+& \left. r_2^{\frac{j}{2}} \frac{\sigma_0 +\sigma_z}{2}\frac{\tau_0+\tau_z}{2} + r_2^{\frac{j}{2}} \frac{\sigma_0 -\sigma_z}{2}\frac{\tau_0-\tau_z}{2} \right\rbrace \otimes |j\rangle \langle j| \nonumber \\
\end{eqnarray}
where $r_2=\frac{\Delta_2+J_2}{\Delta_2-J_2}$. Again, $\mathcal{H}'$  maps to an SSH model of the form
\begin{equation}
    A_2'^{-1} \mathcal{H}' A_2' = \mathrm{i} \sigma_x\left[ \Delta_1\tau_x +\tilde{\Delta}_2 \cos(\hat{k})  \tau_x + \tilde{\Delta}_2 \sin(\hat{k}) \tau_y \right] ,  \label{SSHtrans2}
\end{equation}
where $\tilde{\Delta}_2=\sqrt{\Delta_2^2-J_2^2}$. 

The presence of the NHSE is numerically verified in Fig.~\ref{fig:spcase2}(a), where all eigenstates of $\mathcal{H}'$ are localized at the left edge. As shown in Fig.~\ref{fig:spcase2}(b), nonzero $\omega$ quickly destroys the NHSE, but exactly two localized eigenstates (corresponding to topological modes) are still present. Here, the gap closing conditions are $\tilde{\Delta}_2=\pm \Delta_1$ and $\tilde{\Delta}_2=\pm \mathrm{i} \Delta_1$. Equivalently in terms of the original model parameters, gap closings occur whenever $J_2^2=\Delta_2^2-\Delta_1^2$ or $J_2^2=\Delta_2^2+\Delta_1^2$. In contrast, for PBC, gap closings occur at $J_2^2=(\Delta_1\pm \Delta_2)^2$. These results have also been numerically verified in Fig.~\ref{fig:spcase2}(c,d).

A topological winding number can be constructed in the same way as case ii) with Eq.~(\ref{wn}), but with 
\begin{equation}
    h_\pm = \Delta_1 + \tilde{\Delta}_2 \cos(\hat{k}) \pm \mathrm{i} \tilde{\Delta}_2 \sin(\hat{k}) .
\end{equation}
The topologically nontrivial regime ($w_\pm = \pm 1$) thus corresponds to $|\tilde{\Delta}_2| >\Delta_1 $, whereas the topologically nontrivial regime ($w_\pm = \pm 1$) corresponds to $|\tilde{\Delta}_2| <\Delta_1 $. In terms of the original model parameters, this is equivalent to $J_2 < \sqrt{\Delta_2^2 -\Delta_1^2}$ or $J_2 > \sqrt{\Delta_2^2 +\Delta_1^2}$, which is indeed consistent with the emergence of topological zero modes in Fig.~\ref{fig:spcase2}(c,d).  

\vspace{0.5cm}
\noindent {\it Case iv:} General parameter values
\vspace{0.5cm}

\begin{center}
    \begin{figure}
        \centering
        \includegraphics[scale=0.45]{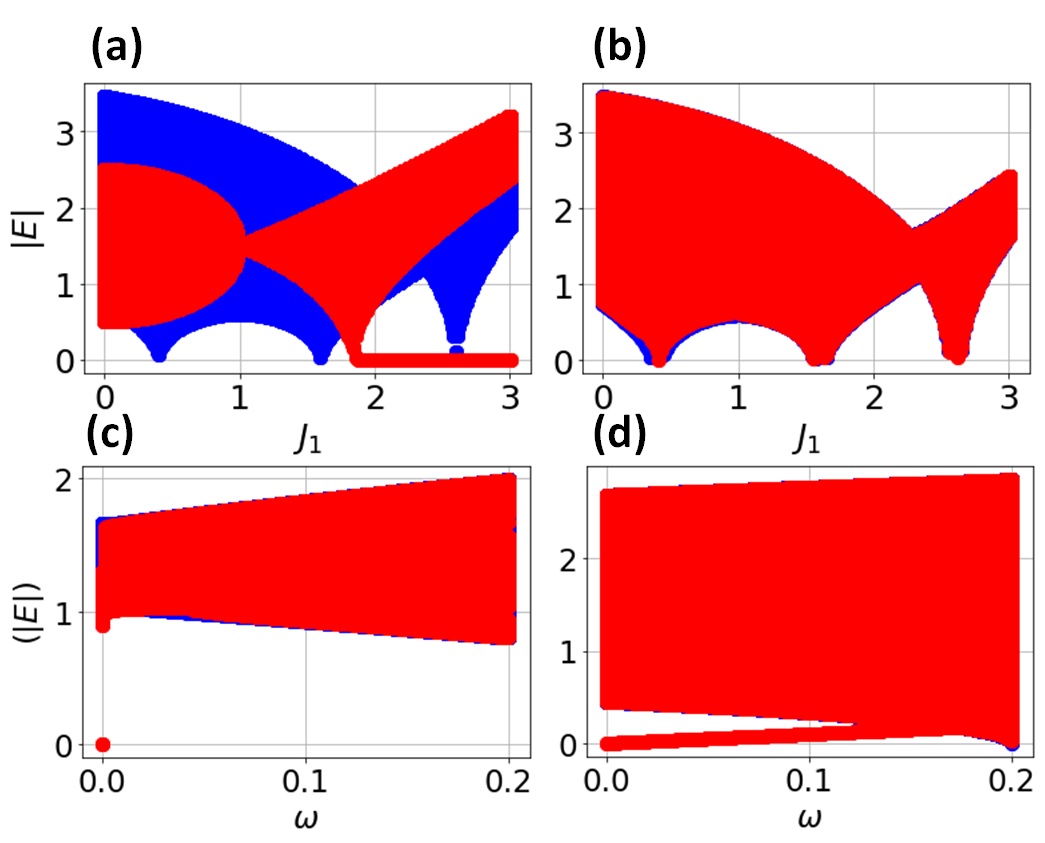}
        \caption{The eigenvalue spectrum of $\mathcal{H}'$ at general parameter values for OBC (red) and PBC (blue). The system parameters are chosen as (a,b,c) $J_2=1$, $\Delta_1=2.1$, $\Delta_2=1.5$, $N=100$, (d) $J_1=1$, $J_2=1.4$, $\Delta_1=1.5$, $\Delta_2=2.1$, $N=100$. The remaining parameters in panels (a,b,c) are (a) $\omega=0$, (b) $\omega=0.05$, (c) $J_1=2.2$. }
        \label{fig:spcase3}
    \end{figure}
\end{center}

We first keep $\omega=0$. In this case,  $\mathcal{A}'=A_1'A_2'$ is a useful mapping that allows us to obtain an effective Hamiltonian
\begin{equation}
    \mathcal{A}'^{-1} \mathcal{H}' \mathcal{A}' = \mathrm{i} \sigma_x\left[ \tilde{\Delta}_1\tau_x +\tilde{\Delta}_2 \cos(\hat{k})  \tau_x + \tilde{\Delta}_2 \sin(\hat{k}) \tau_y \right] . \label{gtrans}
\end{equation}
The gap closing condition is immediately obtained as $|\tilde{\Delta}_2| = |\tilde{\Delta}_1|$. The winding numbers, yet again, are defined via Eq.~(\ref{wn}) with
\begin{equation}
    h_\pm = \tilde{\Delta}_1 + \tilde{\Delta}_2 \cos(\hat{k}) \pm \mathrm{i} \tilde{\Delta}_2 \sin(\hat{k}) .
\end{equation}
The result is nontrivial ($w_\pm =\pm 1$) for $|\tilde{\Delta}_2|>|\tilde{\Delta}_1|$. We recall the relation to original model parameters $\tilde{\Delta}_2=\sqrt{\Delta_2^2-J_2^2}$ and $\tilde{\Delta}_1=\sqrt{\Delta_1^2-J_1^2}$ to make further observations. For instance, it is interesting to note that the topologically nontrivial regime, which supports topological zero modes, exists even when the intracell parameters $J_1$ and $\Delta_1$ are both larger than their intercell counterparts, i.e., $J_2$ and $\Delta_2$ respectively, see e.g., Fig.~\ref{fig:spcase3}(a).  

Next, we consider the case of nonzero $\omega$. Interestingly, here, the transformed excitation Hamiltonian $\mathcal{A}'^{-1} \mathcal{H}' \mathcal{A}'$ acquires an additional term
\begin{eqnarray}
  \mathcal{A}'^{-1} (\omega \sigma_y \tau_0) \mathcal{A}' &=& \omega \sigma_y \tau_0 \mathcal{A}'^2 ,
\end{eqnarray}
which breaks both chiral and translational symmetries. Therefore, in addition to breaking NHSE as we previously demonstrated in Sec.~\ref{NHSE}, such a term is also expected to generally break the system's topology. Indeed, Fig.~\ref{fig:spcase3}(b,c) reveals that the topological zero modes emerging at $J_1>J_2$ and $\Delta_1>\Delta_2$ quickly disappear as soon as nonzero $\omega$ is introduced. However, other topological zero modes, i.e., those that arise when $J_1<J_2$ and $\Delta_1<\Delta_2$, remain present at small $\omega$, but they are no longer pinned at $E=0$ (see \ref{fig:spcase3}(d)). 

In the original bosonic representation, the absence of nontrivial properties at nonzero $\omega$ is attributed to the fact that such an onsite term always leads to a parametric instability, similar to the case of regular BKC \cite{Mcdonald18}. Indeed, the $\sigma_x $ term of $\mathcal{A}'^{-1} \mathcal{H}' \mathcal{A}'$ translates to the number non-conserving $a_j^\dagger a_j^\dagger$ term. Its amplitude increases exponentially with the system size. Fortunately, in a potential experimental setup that we discuss later, realizing our modified Kitaev chain, such a problematic onsite term disappears in a rotating frame for the most natural setup.

\section{Effect of spatial disorder}
\label{disorder}

An attractive feature of a topological phase is its robustness against disorder. It is thus natural to ask if the main features of our modified BKC model, i.e., topological edge states and NHSE, persist in the presence of spatial disorder. To answer this question, we let all parameter values in $H'$ be site-dependent. Specifically, for each lattice site $j$, a parameter $P\in \left\lbrace J_1, J_2, \Delta_1, \Delta_2, \omega \right\rbrace$ is randomly drawn from $[P-W_P P, P+W_P P]$, where $W_P$ is the disorder strength of the parameter $P$.

\begin{center}
    \begin{figure}
        \centering
        \includegraphics[scale=0.45]{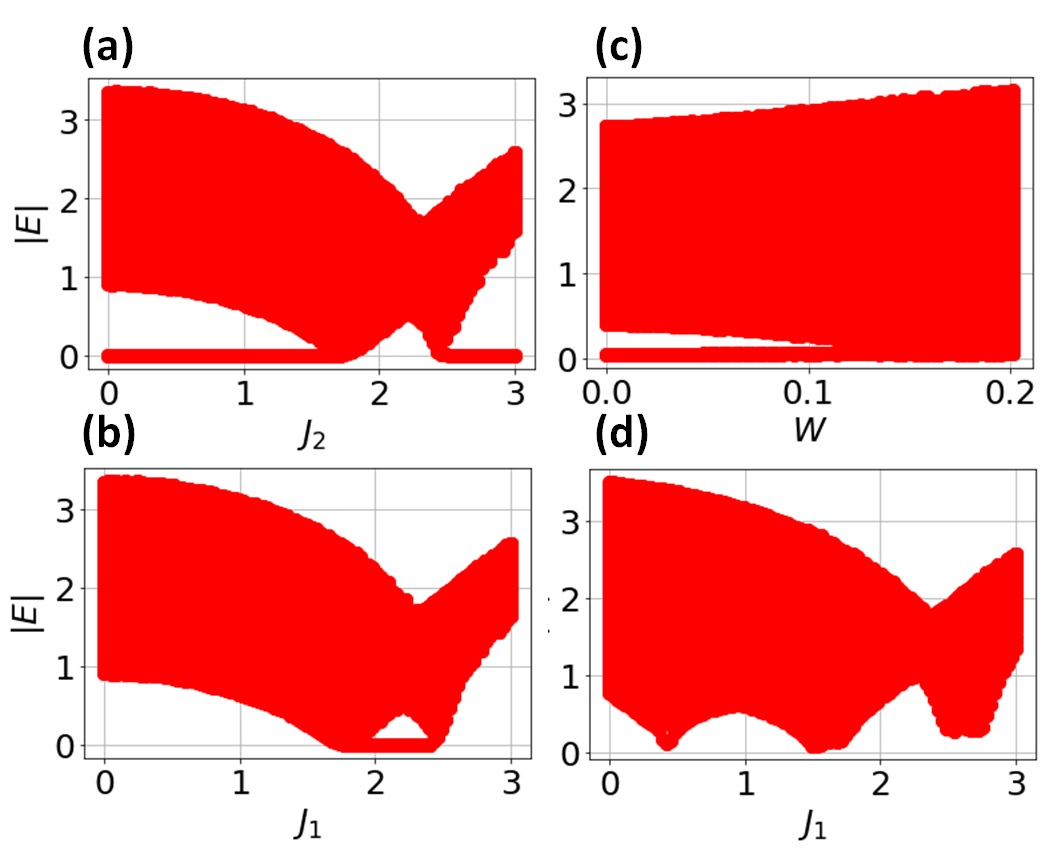}
        \caption{The eigenvalue spectrum of $\mathcal{H}'$ under OBC in the presence of disorder. The system parameters are chosen as (a) $J_1=1$, $\Delta_1=1.5$, $\Delta_2=2.1$, $\omega=0$, (b) $J_2=1$, $\Delta_1=2.1$, $\Delta_2=1.5$, $\omega=0$, (c) $J_1=1$, $J_2=1.4$, $\Delta_1=1.5$, $\Delta_2=2.1$, $\omega=0.05$, (d) $J_2=1$, $\Delta_1=2.1$, $\Delta_2=1.5$, $\omega=0.05$. The disorder strength is set at (a,b,d) $W_P=0.1$ and (c) $W_P=W$ for all parameters. $N=100$ is taken in all panels, and each data point is averaged at over $20$ for disorder realizations.}
        \label{fig:spcase4}
    \end{figure}
\end{center}

We start by setting $\omega=0$. In this case, we could modify the similarity matrices $A_1'$ and $A_2'$ by replacing
\begin{eqnarray}
    r_1^{-\frac{j+1}{2}} \rightarrow  \prod_{\ell\leq j} r_{1,\ell}^{-\frac{1}{2}} &,& r_1^{-\frac{j}{2}} \rightarrow  \prod_{\ell < j} r_{1,\ell}^{-\frac{1}{2}} , \nonumber \\ 
    r_2^{-\frac{j}{2}} & \rightarrow & \left(\prod_{\ell<j} r_{2,\ell}^{-\frac{1}{2}}\right) ,
\end{eqnarray}
where $r_{1,\ell}= \frac{\Delta_{1,\ell}+J_{1,\ell}}{\Delta_{1,\ell}-J_{1,\ell}}$ and $r_{2,\ell}= \frac{\Delta_{2,\ell}+J_{2,\ell}}{\Delta_{2,\ell}-J_{2,\ell}}$, which results in $\mathcal{A}'^{-1}\mathcal{H}'\mathcal{A}'$ taking the form of a disordered SSH model with $\tilde{\Delta}_1 \rightarrow \tilde{\Delta}_{1,j} \equiv \sqrt{\Delta_{1,j}^2-J_{1,j}^2}$ and $\tilde{\Delta}_2 \rightarrow \tilde{\Delta}_{2,j} \equiv \sqrt{\Delta_{2,j}^2-J_{2,j}^2}$. It is thus expected that topological zero modes should remain present as long as the $|E|=0$ gap remains open. This observation is indeed confirmed numerically in Fig.~\ref{fig:spcase4}(a,b) for the two types of zero modes uncovered in the previous section, i.e., those that arise when $J_2>J_1$ and/or $\Delta_2>\Delta_1$ (Fig.~\ref{fig:spcase4}(a)),  as well as those that arise when $J_2<J_1$ and $\Delta_2<\Delta_1$ (Fig.~\ref{fig:spcase4}(b)). Both types of zero modes are equally robust against disorder from all parameters for $\omega=0$.

At nonzero $\omega$, Fig.~\ref{fig:spcase4}(c) reveals that the zero modes that arise when $J_2>J_1$ and/or $\Delta_2>\Delta_1$ continue to be robust against a moderate $\sim 20\%$ disorder in all parameters (including disorder in $\omega$), before the gap at $|E|=0$ closes. However, the zero modes that arise for parameters $J_1<J_1$ and $\Delta_2<\Delta_1$ quickly disappear even for small $\omega\neq 0$ and remain absent in the presence of disorder (see Fig.~\ref{fig:spcase4}(d)). 

\begin{center}
    \begin{figure}
        \centering
        \includegraphics[scale=0.35]{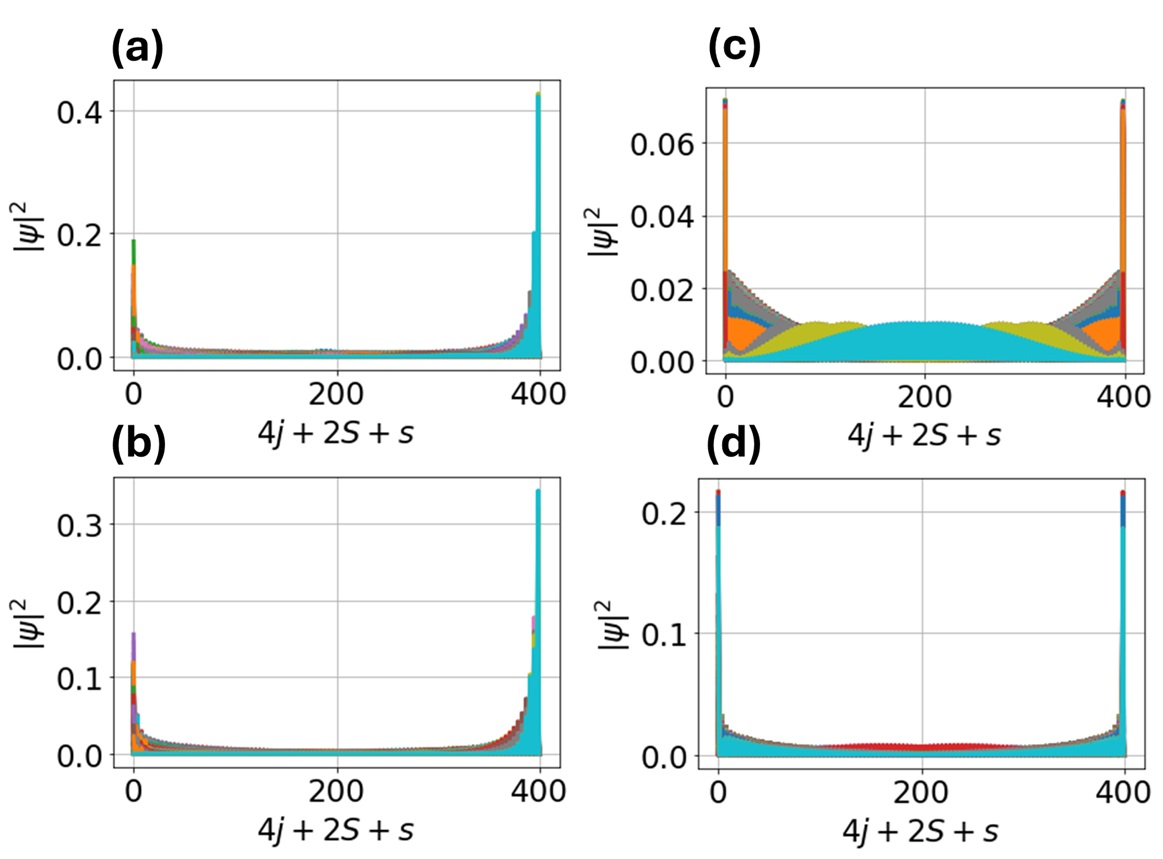}
        \caption{The spatial profiles of all eigenstates of $\mathcal{H}'$ at $N=100$. In panels (a,b), disorder is present in $\omega$ with $W_\omega=2$, and each data point is averaged over $20$ disorder realizations. In panels (c,d), disorder is absent. The system parameters are chosen as (a,c) $J_1=2.2$, $J_2=1$, $\omega=0.05$, $\Delta_1=2.1$, and $\Delta_2=1.5$, (b,d) $J_1=1$, $J_2=0.5$, $\omega=0.05$, $\Delta_1=1.5$, and $\Delta_2=2.1$.}
        \label{fig:spcase5}
    \end{figure}
\end{center}

After uncovering the impact of the disorder on the topological zero modes, we now turn our attention to the fate of the NHSE. To this end, we explicitly plot the spatial profiles of all eigenstates of $\mathcal{H}'$ in Fig.~\ref{fig:spcase5} in two different regimes that support each type of topological zero mode that we obtained above. Interestingly, we observe that, in both regimes, the disorder in $\omega$ allows some features of the NHSE to recover. Indeed, we have verified that a macroscopic number of eigenstates of $\mathcal{H}'$ is localized either to the left or right end of the system at sufficiently large disorder in $\omega$ (Fig.~\ref{fig:spcase5}(a,b)). By contrast, in the absence of disorder, the bulk eigenstates are generically delocalized at finite $\omega$ (Fig.~\ref{fig:spcase5}(c,d)), thereby signifying the absence of NHSE. Any localized solution observed in Fig.~\ref{fig:spcase5}(c,d) is a remnant of the topological zero modes. However, due to broken chiral symmetry, they no longer occur at $|E|=0$.

\section{Possible experimental realization and physical interpretation of instabilities}
\label{sec:exp_real}

Before we conclude, we turn to the question of experimentally realizing the modified BKC model. Such a goal can be slightly tricky - specifically, engineering the number-non-conserving pairing terms and allowing for complex hopping amplitudes requires careful thought. After careful consideration, we provide the following suggested setup that should allow complete control over the parameter space.

We imagine coupled optical cavities embedded in a non-linear medium pumped by external electric fields $E_{\mathrm{pump}}(t)$. Cavity excitation frequencies are assumed to be modulated periodically with a term $\delta_\omega(t)$. The experimental setup is shown below in Fig. \ref{fig:cartoon_setup}.

\begin{figure}[H]
    \centering
    \includegraphics[width=1\linewidth]{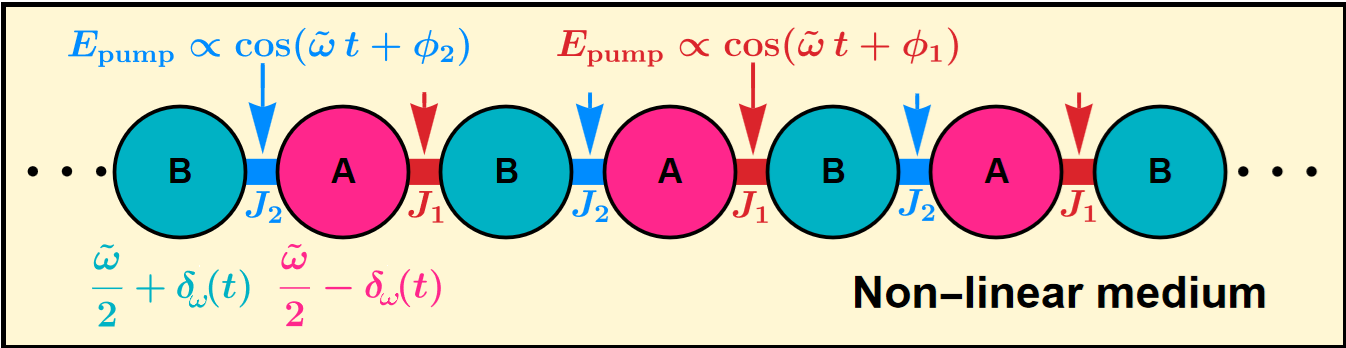}
    \caption{A cartoon of a possible experimental setup of alternating cavities A and B inside of a non-linear medium that has parametrically driven excitation energies $\tilde\omega/2\pm \delta(t)$ and alternating hopping strengths $J_1$ and $J_2$ between cavities. The system is externally driven by an electric pumping field, which is tuned to have alternating phases $\phi_1$ and $\phi_2$ at the bond between cavities.}
    \label{fig:cartoon_setup}
\end{figure}
The Hamiltonian that describes such a system can be taken as
$$H=H_{\mathrm{cav}}+H_{\mathrm{hop}}+H_{\mathrm{e-nl}},$$
where 
$$\scalemath{0.95}{H_{\mathrm{cav}}=\sum_j\left[\left(\frac{\tilde\omega}{2}+\delta_\omega(t)\right) a_{A, j}^{\dagger} a_{A, j}+\left(\frac{\tilde\omega}{2}-\delta_\omega(t)\right) a_{B, j}^{\dagger} a_{B, j}\right]}$$
describes cavity modes that are modulated periodically with period $T$ (for instance, by a vibrating mirror) according to 
$$\delta_\omega(t)=\lambda\frac{\pi ^2 \sin \left(\frac{2 \pi  t}{T}\right)}{2 T}.$$
We will see later that parameter $\lambda$ (it is zero if the cavity is not parametrically driven) allows us to tune if parameter $J_i$ in the final model is real or complex. Particularly, we will see that $\lambda=0$ corresponds to purely real $J_i$, $\lambda=1$ to purely imaginary $J_i$, whereas real values between $0$ and $1$ yield complex $J_i$. Next, we allow for tunneling between cavities that are described by real-valued hopping parameters $\tilde J_i$ via the Hamiltonian
$$H_{\mathrm{hop}}=\sum_j\left[\tilde J_1 a_{A, j}^{\dagger} a_{B, j}+\tilde J_2 a_{B, j}^{\dagger} a_{A, j+1}+h.c.\right].$$
Finally, we model an interaction between cavities, non-linear medium, and pump fields in a typical non-linear response form by a term of the form $\propto \chi^{(2)} E_{\mathrm{pump }}(t) E_i E_{i+1}$, where $\chi^{(2)}$ is second order susceptibility. For our cavity modes and choice of oscillatory external field, we combine constants into $\tilde\Delta_i$ and find an expression of the form
\begin{equation}
\begin{aligned}
&H_{\mathrm{e-nl}}= \sum_j\left[\tilde \Delta_1 \cos \left(\tilde\omega t+\phi_1\right)\left(a_{A, j}+a_{A, j}^{\dagger}\right)\left(a_{B, j}+a_{B, j}^{\dagger}\right)\right. \\
& +\left.\tilde \Delta_2 \cos \left(\tilde\omega t+\phi_2\right)\left(a_{B, j}+a_{B, j}^{\dagger}\right)\left(a_{A, j+1}+a_{A, j+1}^{\dagger}\right)\right].
\end{aligned}
\end{equation}
We may now use a rotating frame transformation 
$$U=\exp\left(-i\int dt H_{\mathrm{cav}}\right)$$
and assume that frequency $\tilde \omega$ is large enough to enact a subsequent rotating wave approximation - in our case, just a time average over $2\pi/\tilde\omega$. Moreover, we assume in particular that $\tilde\omega \gg \frac{2\pi}{T}$  which ensures that $\delta_\omega(t)$ is unaffected by time averaging over time $2\pi/\tilde\omega$. We then obtain the effective Hamiltonian
\begin{equation}
    H_{\mathrm{eff}}=H_{\mathrm{eff,hop}}+H_{\mathrm{eff,e-nl}},
\end{equation}
where 
\begin{eqnarray}
H_{\mathrm{eff,hop}}&=&\sum_j\left[
\tilde J_1 e^{i\chi(t)} a_{A, j}^{\dagger} a_{B, j} +\tilde J_2 e^{-i\chi(t)} a_{B, j}^{\dagger} a_{A, j+1}\right. \nonumber \\
\left. +h.c.\right]    
\end{eqnarray}
is a modified hopping term and we defined $\chi(t)=2\int_0^t dt^\prime\delta_\omega(t^\prime)$. We also find the effective boson pairing term 
$$H_{\mathrm{eff,e-nl}}=\sum_j\left[ e^{i\phi_1}\tilde\Delta_1 a_{A, j}^{\dagger} a_{B, j}^{\dagger}+\tilde\Delta_2 e^{i\phi_2}a_{B, j}^{\dagger} a_{A, j+1}^{\dagger}+h.c.\right]$$
which is just the effective number-non-conserving term we wanted. We realize that for the case of $\lambda=0$ (cavity energies are not parametrically driven), we have achieved a version of the Hamiltonian $H^\prime$ that has real-valued $J_i=\tilde J_i$ and $\Delta_i=e^{i\phi_i}\tilde\Delta_i$, which can be tuned to any phase.

Note that when $\lambda \neq 0$, the resulting hopping part of the Hamiltonian ($H_{\mathrm{eff,hop}}$) is still time-dependent. In this case, we shall further assume that $\frac{2\pi}{T}$, despite being much smaller than $\tilde{\omega}$, is still large compared to other remaining model parameters, such that another time average (this time with respect to $T$) can be applied. Note that this decoupling of scales justifies the independence of the two time averaging processes. Following a typical approach from Floquet theory, we can now take the second time average to obtain a final effective time-independent Hamiltonian. Specifically in the case $\lambda=1$, we find that $\frac{1}{T}\int_0^T dt e^{\pm i\chi(t)}=\pm iJ_0(\pi/2)$ with $J_0$ the typical Bessel function of the first kind, i.e., we managed to engineer imaginary values for the hoppings. With different values for $\lambda$, we can, in principle, engineer any kind of complex phase (see Eq.~(\ref{eq:mod_hopp}) below).

This result now allows us to identify terms with the Hamiltonian $H^\prime$. We have
\begin{equation}
   \scalemath{0.95}{ J_1=e^{i\frac{\pi\lambda}{2}}J_0\left(\frac{\pi\lambda}{2}\right)\tilde J_1;\; J_2=e^{-i\frac{\pi\lambda}{2}}J_0\left(\frac{\pi\lambda}{2}\right)\tilde J_2;\; \Delta_i=e^{i\phi_i}\tilde \Delta_i};
   \label{eq:mod_hopp}
\end{equation}
and $\omega=0$ for the case of general $\lambda$, which shows clearly that our setup allows to tune any complex values for $J_i$. We note that the resultant Hamiltonian, as we have discussed here, does not include the onsite term ($\omega$). Given that the presence of such an onsite term is detrimental to the formation of NHSE and some topological edge states in our modified BKC, the fact that our proposed experimental setup does not naturally yield the $\omega$ term further demonstrates the relevance of the nonHermitian physics we uncovered for our model. Nevertheless, if needed, it is also possible to engineer an additional onsite term in the system by making our cavity excitation energies $\tilde\omega/2$ in $H_{\mathrm{cav}}$ differ slightly from the term $\tilde\omega$ that appears in the time dependence due to the electric pump field in $H_{\mathrm{e-nl}}$ and otherwise following the same procedure.

It is also interesting to notice that at precisely $\frac{2\pi}{T}=n\tilde\omega$ with $n\in\mathbb{N}$, we directly obtain Eq.~(\ref{eq:mod_hopp}), i.e. the hopping terms that are tunable via $\lambda$, under rotating wave approximation only - no additional time average and assumptions needed. We chose to center the discussion around employing the hierarchy ($\tilde\omega\gg \frac{2\pi}{T}\gg$ other parameters) so we could expose additional flexibility in the choice of parameters for an experimental realization of the model that requires no fine-tuned relation between $\tilde\omega$ and $T$. Therefore, depending on which is easier to achieve in actual experiments, we could realize the complex hoppings of Eq.~(\ref{eq:mod_hopp}) either without an additional time averaging process at the expense of fine tuning $T$ and $\tilde{\omega}$, or at more general values of $T$ and $\tilde{\omega}$ but at the expense of requiring an additional time averaging process. Either way, our proposed experimental setup offers total control also over the phase of hopping parameters via choice of parametric driving strength $\lambda$.

Lastly, we comment that the modified BKC Hamiltonian $H^\prime$ in a generic parameter regime actually cannot be paraunitarily diagonalized, i.e., a means of diagonalizing the system's second quantized Hamiltonian in terms of purely bosonic modes \cite{Colpa}. This is commonly interpreted to signify an instability, as the breakdown of a paraunitary diagonalization procedure can be interpreted to signify problems with a bosonic ground state \cite{nobuyuki}. For instance, such a breakdown happens if there exist negative excitation energies, which would lead to an infinitely occupied state, which corresponds to an unstable ground state. In our case, however, this is not a problem because our experimental proposal involves the use of periodic driving. In such a driven system, the concept of a ground state is not properly defined anyway. Moreover, as this work focuses on the topological and the effective nonHermitian properties of the system, we do not strictly require the energy excitations to be purely bosonic in nature.

\section{Concluding remarks}
\label{conc}

In this paper, we devised a modified BKC, the excitation Hamiltonian of which supports nontrivial topological phases and NHSE. We analytically constructed an exact similarity transformation that maps our model into two copies of nonHermitian SSH models, enabling a complete topological characterization. In particular, two types of topological edge modes were identified; one occurs when an intercell parameter (hopping or pairing strength) is dominant over its intracell counterpart, and the other when both intracell parameters are dominant over their intercell counterparts. In addition to verifying the presence of topological edge modes, we also numerically confirmed the formation of the NHSE, marked by the sensitivity of the excitation energy spectrum against the choice of the boundary conditions, as well as the localization of all bulk excitation eigenstates at system edges.

The presence of an onsite (harmonic oscillator) term, no matter how small, was found to destroy the NHSE, thus delocalizing all bulk excitation eigenstates. However, some topological edge modes survive upon introducing such an onsite term, while others disappear quickly. We have further verified that all topological edge modes are robust against considerable spatial disorder. Moreover, the presence of spatial disorder was found to stabilize the NHSE, i.e., a macroscopic number of localized eigenstates are preserved even at nonzero bosonic frequency. We also discussed a possible setup that might experimentally realize the model and would give total control over the parameter space.

As a potential future work, further modifications to the BKC could be envisioned, and we expect them to yield an even richer interplay between topology and the NHSE. For example, one might consider time-periodic system parameters \cite{Bom25}. The resulting system is expected to generate an effective Floquet nonHermitian excitation Hamiltonian, which potentially supports novel topological effects with no static counterparts \cite{Zhou18,Zhou21,Zhou22,Wu20,Cao21,Wu21,Zhang20b,Vyas21,Weidemann22}. Alternatively, by introducing nonHermiticity into the underlying bosonic Hamiltonian itself, e.g., by making either the hopping or bosonic pairing terms nonreciprocal in the second quantized Hamiltonian, the excitation Hamiltonian may contain a different form of nonHermiticity not achievable here, e.g., onsite non-Hermiticity (gain/loss). Finally, combining the BKC with an extended SSH \cite{Alvarez19,Anastasiadis22,Du24,Ghuneim24} instead of the regular SSH model considered in the present work may result in additional topological edge modes beyond zero modes.

\section*{Acknowledgments}
M.V. and R.W.B. gratefully acknowledge the support provided by the Deanship of Research Oversight and Coordination (DROC) and the Interdisciplinary Research Center(IRC) for Intelligent Secure Systems (ISS) at King Fahd University of Petroleum \& Minerals (KFUPM) for funding their contribution to this work through internal research grant No. INSS2507. J.L. acknowledges the support of the Natural Sciences and Engineering Research Council of Canada (NSERC) RGPIN-2022-03882 and (NRC) AQC-200-1.

\section*{Code and data availability}
The data that supports the findings of this study are available from the corresponding author upon request.

\end{document}